\documentclass{article}
\usepackage{graphicx}
\usepackage{amsmath}

\input{tcilatex}

\begin{document}

\begin{center}
{\Large MAGNETIC FRACTAL\ DIMENSIONALITY OF THE DIELECTRIC\ BREAKDOWN UNDER
STRONG MAGNETIC FIELDS\ }
\end{center}

\bigskip

\begin{center}
\textit{Y. Ben-Ezra}$^{1,2,3}$\textit{\ , Yurij V. Pershin}$^{4,2}$\textit{,
I.D.Vagner}$^{1,2}$\textit{\ and P.Wyder}$^{2}$\textit{\ }

\textit{\bigskip }

$^{1}$\textit{P.E.R.I.-Physics and Engineering Research Institute at Ruppin, 
\\[0pt]
Emek Hefer, 40250 \ Israel. }

$^{2}$\textit{Grenoble High Magnetic Fields Laboratory \\[0pt]
Max-Planck-Institute f\"{u}r Festkorperforschung and CNRS,\\[0pt]
166X, F-38042, Cedex 9, France.\\[0pt]
}

$^{3}$\textit{School of Physics and Astronomy, Tel Aviv University, \\[0pt]
Tel Aviv 69978, Israel.\\[0pt]
}

$^{4}$\textit{B.I.Verkin Institute for Low Temperature Physics and
Engineering, }

\textit{47 Lenin Avenue 310164 Kharkov, Ukraine. \\[0pt]
}

\bigskip
\end{center}

The formation of breakdown pattern on an insulating surface under the
influence of a transverse magnetic field is theoretically investigated. We
have generalized the Dielectric Breakdown Model (DBM) for the case of
external magnetic field. Concept of the Magnetic Fractal Dimensionality
(MFD) is introduced and its universality is demonstrated. It is shown that
MFD saturates with magnetic fields. The magnetic field dependence of the
streamer curvature is obtained. It is conjectured that nonlinear current
interaction is responsible for the experimentally observed 'spider-legs'
like streamer patterns.\hspace{1in}

\bigskip

PACS numbers: 52.20.Dq, 52.80.Mg

\bigskip

\newpage

Fractal properties are common to the dielectric breakdown phenomena \cite
{Erzan} which range from an atmospheric lightning \cite{lichtenberg},\cite
{vols}, \cite{wyder} to electric treeing in polymers \cite{fava} and are of
scientific and technical importance \cite{Zeller},\cite{Vicsek}. Although
the actual physical processes can be quite different in these phenomena, the
global properties of the resulting discharge patterns are very similar.
Filamentary gas discharges on insulating surface exhibit remarkable
similarities to\ breakdown phenomena in long gaps \cite{Gallimberti}, e.g.,
to atmospheric lightning , and thus offer the possibility to perform
well-defined model experiments in laboratory. The application of
sophisticated diagnostic tools, such as streak cameras, high-speed
oscilloscopes, and time-resolved spectroscopy \cite{Larigadle}, in surface
discharge experiments has improved the basic understanding of how a highly
conducting phase, the filamentary ''leader'' channel, advances into a
nonconducting \ medium as the surrounding gas.

The surface discharge in compressed $SF_{6}$ gas have been studied in detail
by Niemeyer and Pinnekamp\cite{Pinn}. The parameters were controlled in such
a way that the experiment produces, to a good approximation, an
equipotential channel system growing in a plane with a radial electrode from
a central point. The experiment shows that the dielectric breakdown pattern
has a fractal structure.

\qquad The stochastic model containing the essential features of the fractal
properties of the dielectric breakdown was introduced by Niemeyer,
Pietroniero, and Wiesmann (NPW) \cite{NPW},\cite{Erzan}. The introduced
stochastic model made it possible to simulate the growth of a fractal
structure which resembles a Lichtenberg figure \cite{lichtenberg}. The basic
assumption of the dielectric breakdown model (DBM) was that the growth
probability depends on the local field. This model naturally leads to
fractal structures. In \cite{NPW} the Haussdorf dimension and other fractal
properties of the structure are determined and their close relations to the
other fractal structures e.g., DLA (the diffusion limited aggregation) \cite
{witten}, is shown .

\qquad The DBM\ method was generalized by Wiesmann and Zeller \ \cite
{wiesm86} in two ways. Firstly they introduced a critical field for growth $%
F_{c}$. The growth probability was assumed to be proportional to the local
field $F_{loc}$ if $F_{loc}>F_{c}$ and zero if $F_{loc}<F_{c}.$ Secondly
they have introduced an internal field in the structure $F_{s}$. The
potential in the structure was no longer equal to the potential $V_{0}$ of
the connecting electrode but equal to $V_{0}+F_{s}\cdot s$, where $s$ is the
length of the path along the structure which connects the point to the
electrode.

\qquad In the recent experiment \ \cite{wyder} a transverse high magnetic
field was applied during the discharge evolution and thus any spatial
restriction of the surface discharge was avoided in order to use a locally
sensitive probe.

When a rectangular high-voltage pulse of about $500ns$ length, $20kV$
amplitude, and less than $10ns$ risetime is applied to a point-to- plane
electrode system, the electrodes being separated by a thin dielectric film (
thickness $\approx 100\mu m$) , a discharge propagates in the gas just above
the surface of the film. The discharge pattern was recorded directly with
high resolution by using a photographic film as dielectric plate.

The experiment shows the spatial evolution of a negative surface discharge
in a nitrogen atmosphere as a function of the magnetic field ( Fig.2 of Ref. 
\cite{wyder}). At $B=0$ a very bright starlike pattern develops. At moderate
magnetic field (up to $7T$ ), the leader channels are bent and appear to
have a circular shape outside the central electrode region.

The radius of curvature is of the order of $1cm$ at $7T$. The direction of
the bending corresponds to the movement of electrons in crossed \ electric
and magnetic fields. With the increase of the magnetic field the radius of
curvature decreases, the channels approach each other and branching sets in.
At the highest applied magnetic field of $12T$, circular-shaped current
filaments are only found in the outmost regions of the discharge pattern
where they can develop undistributed by the fields of neighboring leader
channels.

One can summarize the experimental results as follows:

(1) The observed bending effect cannot be related to the movement of a
single charged particles in crossed electric and magnetic fields, which
should result in the curvatures of the order of magnitude of $15\mu m$ \ at $%
1T$ \ for the electrons, i.e., the Larmor radius $r_{c}=v_{c}/\omega _{c}$
of the cyclotron orbit of an electron withe enough kinetic energy to ionize
gas molecules. It has be assumed that the macroscopic value of the radius of
curvature results from the drift movement of the Larmor centers of the
gyrating electrons. The average radius of curvature follows a power law, \ $%
R\propto B^{-\alpha }$ . The exponent $\alpha $ depends on the nature of gas.

(2) The circular shape of the current filaments is not expected to occur for
our point-to-plane electrode geometry.

(3) The complexity of the discharge pattern can be controlled by the
magnetic field strength. It appears that an increase in complexity is
induced on the scale of the bending observed, and not on the microscopic
scale of the cyclotron radius.

In what follows we modify the model introduced by Wiesmann and Zeller \cite
{wiesm86} for the case of external magnetic field.

Let us consider a two-dimensional square lattice in which a central point
represents one of the electrodes while the other electrode is modeled as a
circle at large distance, which represents the geometry of the experiment.
The discharge starts at the central electrode and grows by one lattice bond
per growth step. A bond connects a lattice point with one of eight adjacent
lattice points as it is described in Fig.1. Once a given point is connected
to the discharge structure by a bond, it becomes part of the structure. The
potential of the central electrode is $0$, the potential of a point in the
structure is $V_{i,k}=V_{l,m}+V_{R}\ast l$, where $V_{l,m}$ is the potential
of a points from which growth go on, $V_{R}$ is an internal field in the
structure and $l$ is $1$ for bonds parallel to the grid and is $\sqrt{2}$
for diagonal bonds.

The growth is computed as follows: First the Laplace equation is solved with
the boundary conditions determined by the electrodes and the discharge
structure. Then the local field $F_{loc}$ between a point which is already a
part of the structure $\left( i,k\right) $ and a new adjacent point $\left(
i^{\prime },k^{\prime }\right) $ is calculated:

\begin{equation}
F_{loc}(i,k,i^{\prime },k^{\prime })=\frac{\varphi _{_{i^{\prime },k^{\prime
}}}-V_{i,k}}{l}
\end{equation}
where $\varphi _{_{i^{\prime },k^{\prime }}}$ is the solution of Laplace
equation. The breakdown can occur only if the local field is greater that \
the critical field of growth $F_{C}$. The probability that a new bond will
form between a point which is already a part of the structure and a new
adjacent point $p(i,k\rightarrow i^{\prime },k^{\prime })$ is calculated as
a function of the local field $F_{loc}$ between the two points:

\begin{equation}
p(i,k\rightarrow i^{\prime },k^{\prime })=\left\{ 
\begin{array}{ccc}
0 & , & F_{loc}(i,k,i^{\prime },k^{\prime })<F_{C} \\ 
&  &  \\ 
\frac{F_{loc}^{\eta }}{\sum F_{loc}^{\eta }} & , & F_{loc}(i,k,i^{\prime
},k^{\prime })\geq F_{C}
\end{array}
\right.  \label{probab}
\end{equation}
where a power-low dependence with exponent $\eta $ is assumed to describe
adequately the relation between the local field and the probability. The sum
in the denominator refers to all possible growth processes. A new bond is
chosen randomly with probability distribution (\ref{probab}) and added to
the discharge pattern. With this new discharge pattern one starts again.
More detailed description of this model is done in \cite{wiesm86} .

Let us take into account the magnetic field. The magnetic field changes the
probability distribution (\ref{probab}). A moving particle in the magnetic
field experiences the Lorentz force $F_{L}\sim \overrightarrow{V}\times 
\overrightarrow{H}$ which acts perpendicularly to its velocity. Consider
each step of growth like a superposition of two processes. The first step is
choosing a new bond using the probability distribution (\ref{probab}), and
the second step is taking into account the probability of deviation of the
growth due to the magnetic field, $p_{H}$.

If after first step of growth, for example, the bond from the dot $0$ to the
dot $4$ was chousen (Fig.1), the growth will occurs to the dot $3$ with the
probability $p_{H}$ and to the dot $4$ with the probability $1-p_{H}$. The
new probability of growth can be written as

\begin{equation}
\widetilde{p}(i,k\rightarrow i^{\prime },k^{\prime })=\frac{p(i,k\rightarrow
i^{\prime },k^{\prime })(1-p_{H}(i,k,i^{\prime },k^{\prime
}))+p(i,k\rightarrow i^{\prime \prime },k^{\prime \prime
})p_{H}(i,k,i^{\prime \prime },k^{\prime \prime })}{\sum (p(i,k\rightarrow
i^{\prime },k^{\prime })(1-p_{H}(i,k,i^{\prime },k^{\prime
}))+p(i,k\rightarrow i^{\prime \prime },k^{\prime \prime
})p_{H}(i,k,i^{\prime \prime },k^{\prime \prime }))}  \label{newprob}
\end{equation}
where the point $\left( i^{\prime \prime },k^{\prime \prime }\right) $ is
the neighboring point with respect to the point $\left( i^{\prime
},k^{\prime }\right) $ in the clockwise direction with respect to the point $%
\left( i,k\right) $ and the sum in the denominator refers to all possible
growth processes. The probability \ref{newprob} was used in computer
simulations in place of the probability \ref{probab} with the same algorithm.

Let us turn to the probability $p_{H}$ of deviation of the growth due to the
magnetic field. In our model, during the process of growth, two constant
forces act on charge cariers, $F_{L}$ and $F_{H}$ (Fig.1). When the
resulting force is near the dot $3$, $p_{H}\longrightarrow 1$; when the
resulting force is near the dot $4$, $p_{H}\longrightarrow 0$. \ It is clear
that the probability $p_{H}$ is proportional to $\frac{F_{L}}{F_{loc}}$ . If 
$F_{loc}(i,k,i^{\prime },k^{\prime })<F_{C}$ then the growth does not occur
and $p_{H}$ should be zero. If $\frac{F_{L}}{F_{loc}}>1$ then we let $%
p_{H}=1 $. The Lorentz force $F_{L}$ is proportional to the velocity of
charge carrier. In our model we can take into account only a local velocity,
which arrises in the first step of growth process due to the acceleration in
the local field $F_{loc}$. So, the Lorentz force should be proportional to $%
\sqrt{lF_{loc}(i,k,i^{\prime },k^{\prime })}$.

Based on these considerations, we choose the probability of deviation of the
growth due to the magnetic field, $p_{H},$ in the following form:

\begin{equation}
p_{H}(i,k,i^{\prime },k^{\prime })=\left\{ 
\begin{array}{ccc}
0 & , & F_{loc}(i,k,i^{\prime },k^{\prime })<F_{C} \\ 
&  &  \\ 
\frac{F_{L}(i,k,i^{\prime },k^{\prime })}{F_{loc}(i,k,i^{\prime },k^{\prime
})} & , & \frac{F_{L}(i,k,i^{\prime },k^{\prime })}{F_{loc}(i,k,i^{\prime
},k^{\prime })}\leq 1,\left. {}\right. F_{loc}(i,k,i^{\prime },k^{\prime
})\geq F_{C} \\ 
&  &  \\ 
1 & , & \frac{F_{L}(i,k,i^{\prime },k^{\prime })}{F_{loc}(i,k,i^{\prime
},k^{\prime })}>1,\left. {}\right. F_{loc}(i,k,i^{\prime },k^{\prime })\geq
F_{C}
\end{array}
\right.
\end{equation}

where $F_{L}(i,k,i^{\prime },k^{\prime })=\sqrt{lF_{loc}(i,k,i^{\prime
},k^{\prime })}H$ and $H$ is the value of the magnetic field.

In our computer simulations we consider a $500\times 500$ lattice. The
solutions of the Laplace equation were obtained by the iteration method \cite
{NPW}. Before starting of each realization of growth we performed $20000$
iterations and after each step of growth the number of iterations was $40$.
This procedure gives a good convergence. The number of particles in clusters
was $9000$.

The fractal dimension was calculated by the method described in \cite{NPW}.
For every realization we plotted the $\log N\left( R\right) $ versus $\log R$
where $N\left( R\right) $ is the number of particles belonging to the
structure and being within a \ circle of radius $R$. The fractal dimension
is obtained by fitting a straight line to the data scaling region. For every
set of the same parameters of the model ($H$, $F_{C}$, $V_{R}$, $\eta $) we
made about $100$ realizations. Thus the statistical fluctuations were
reduced.

Let us discuss the results of these calculations.

We start our simulations with the case of the zero magnetic field and the
zero values of the parameters $Fc$ and $V_{R}$ in order to compare our
results with the results of the different authors. In the Table.1 we present
the dependence of \ the Hausdorff dimension $D$ on the exponent parameter $%
\eta .$

\bigskip

\begin{tabular}{|c|c|c|c|}
\hline
$\eta $ & $D$ (our results) & $D$ (according to\ \cite{NPW}) & $D$
(according to\ \cite{wiesm86}) \\ \hline
$0.5$ & $1.89\pm 0.02$ & $1.89\pm 0.01$ & $1.92$ \\ \hline
$1$ & $1.73\pm 0.02$ & $1.75\pm 0.02$ & $1.70$ \\ \hline
$2$ & $1.6\pm 0.03$ & $\sim 1.6$ & $1.43$ \\ \hline
\end{tabular}

\bigskip Table 1. Dependence of the Hausdorff dimension $D$ on the exponent $%
\eta $ used in the relation between probability and local field (Eq.\ref
{probab}).

\bigskip

Our results are in a good agreement with the results, obtained earlier.

The example of the computer-generated discharge pattern (Lichtenberg figure 
\cite{lichtenberg}) corresponding to the following set of parameters: $%
H=0,\eta =1,F_{c}=0$ and $V_{R}=0$ is shown in the Fig.2. In the Fig.3 \ we
show the computer-generated discharge pattern in the presence of the
external magnetic field. The white lines in the figures 1 and 2 correspond
to the leader channels. Unlike the pattern, presented in the Fig.2, the
leader channels in the magnetic field are distorted and appeare to have a
circular shape outside the central electrode region. The direction of the
bending corresponds to the movement of electrons in crossed electric and
magnetic field. At the lower left corner of the Fig.3 we plotted the
cyclotron orbit of an electron. One can note that the Larmor radius of an
electron is about two orders of magnitude smaller then the radius of
curvature of the leader channel.

The saturation of the fractal dimensionality, with growing magnetic fields,
at the value of $D=1.67$ , which is one of the main results of this paper,
is presented in Fig. 4. \ The plot starts from the value of $D=1.65$ , in
the absence of magnetic field, which is smaller than the fractal
dimensionality reported in \cite{NPW}. Such difference results from the fact
that in \cite{NPW} the critical field value for the breakdown was not taken
into account. We have improved their calculations by introducing the minimal
value of the electric field for the breakdown between two successive points.
In this case the breakdown pattern is more directionally selected, and a
lower fractal dimensionality results. With the increase of the external
magnetic field, the MFD growth and finally saturates at an universal for
high magnetic fields values of $1.67\pm 0.01.$ The growth of MFD with the
magnetic field could be expected, since the curved trajectories fill up the
space more densely than the straight ones. The existence of an universal
limit, hovewer, is far from being obvious. Following the directed
percolation models, one could think that the saturation of MFD will occur at 
$D=2$ . Fig 4. shows clearly that in this system the MFD saturates due to
the physics of current carrying streamers.

An unexplained feature in the experiments of \cite{wyder} is the
'spider-legs' form of the breakdown pattern in the absence of the external
magnetic field. \ We outline here that the streamer currents are rather
strong, $10\div 100$ $A$, \ and their influence on the streamer pattern can
be very important. To describe this phenomenon we have taken into account
the magnetic interaction between current carrying streamers, in the
framework of the modified active walker model (MAWM) which is described in
what follows. The results of our calculations, presented in Fig. 5, show
that our model correctly describes the experimentally observed 'spider-legs'
effect.

Active walker models have been used to describe different pattern formation
problems \cite{Chia99,HSKM97}. In these models the walkers moovement is
subject to the influences of the environment and vice versa. We describe the
leader channel propagation in the magnetic field using the active walker
model. The Lorentz force acting on the fast-moving electrons is particularly
effective in the high-field regions in the leader tips, where the channel
formation takes place.

Let us consider again a two-dimensional square lattice in which a central
point represents one of the electrodes while the other electrode is modeled
as a circle at large distance. The discharge starts at the central
electrode, so initially several walkers are set in the vicinity of it. The
walkers move in a potential which is the solution of the Laplace equation
with the boundary conditions determined by electrodes and discharge
structure. During a step of growth each walker moves. The solution of
Laplace equation is found by iteration method \cite{NPW} after each step of
growth.

When a walker moves to a point, this point starts belonging to the breakdown
structure. The potential of the central electrode is $0$, the potential of a
point in the structure is $V_{i,k}=V_{l,m}+V_{R}\ast l$, where $V_{l,m}$ is
the potential of a points from which growth go on, $V_{R}$ is an internal
field in the structure and $l$ is $1$ for bonds parallel to the grid and is $%
\sqrt{2}$ for diagonal bonds. In the absence of the magnetic field, the
probability of a walker step is a function of the local field $F_{loc}$ \cite
{Chia99}. The breakdown occurs only if the local field is greater then \ the
critical field of growth $F_{C}$.

The magnetic field is taken into account by the following way. We add to the
local field in the direction perpendicular to the previous move of the
walker the Lorentz force which is proportional to the magnetic field. The
magnetic field acting on the $i$ walker is $H_{i}=H_{0}+H_{I}^{i}$, where $%
H_{0}$ is external magnetic field and $H_{I}^{i}$ is the field created by
currents of the breakdown structure. We calculate $H_{I}^{i}$ by means of
the Biot-Savart law:

\begin{equation}
H_{I}^{i}=\sum\limits_{k\neq i,l}\frac{I}{r^{3}}d\overrightarrow{s}%
_{k,l}\times \overrightarrow{r}
\end{equation}

where $I$ is the current in a channel, $d\overrightarrow{s}_{k,l}$ is $l$
element of the $k$ channel and $\overrightarrow{r}$ is the vector pointed
from the position of $d\overrightarrow{s}_{k,l}$ to the position of the $i$
walker.

The main results of this part of studies are presented in Fig. 5. \ 

To summarize, we have generalized and modified the existing dielectric
breakdown models to explain the experimental observations of the propagation
of a streamer near an insulating surface under the influence of a transverse
magnetic field. We have introduced the concept of the Magnetic Fractal
Dimensionality (MFD) and have obtained its saturation with growing magnetic
fields. The Universal Magnetic Fractal Dimensionality (UMFD) equals $1.670$
which is superior to $1.65,$ the one in the absence of a magnetic field. \
Inclusion of the magnetic interaction between the current-carrying streamers
results in the 'spider-legs' like streamer patterns at lower fields, which
corresponds to the experimental observations.

We acknowledge helpfull discussions with T.Maniv, and A.Zhuravlev. We
acknowledge the assistance of A.Kaplunovsky in the initial stage of
numerical simulations.

\textbf{\newpage }

\bigskip Figure Captions

\bigskip 

Fig.1.

Bonds connecting a lattice point with one of eight adjacent lattice points
in a two-dimensional square lattice. The geometry of the experiment is
represented by a central electrode at some point while the other electrode
is a circle at large distance. The discharge starts at the central electrode
and grows by one lattice bond per growth step.

\bigskip

Fig.2.

Computer-generated discharge pattern (Lichtenberg figure) corresponding to
the following set of parameters: $H=0,\eta =1,F_{c}=0$ and $V_{R}=0$ as
explained in the text.The white lines correspond to the leader channels.

\bigskip

Fig.2.

Computer-generated discharge pattern in the presence of the external
magnetic field. The leader channels in the magnetic field are distorted and
appeare to have a circular shape outside the central electrode region, due
to the action of the magnetic field.

\bigskip

Fig.4.

The saturation of the fractal dimensionality, with growing magnetic fields,
at the value of $D=1.67$ .

\bigskip

Fig.5.

The 'spider-legs' form of the breakdown pattern folowing from the magnetic
interaction between the streamer currents: a) current-current interactions
are not took into account, $H=0$; b)current-current interactions are took
into account, $H=0$; c) $H\neq 0$.

\end{document}